\documentclass[showpacs, twocolumn, amsmath, amssymb]{revtex4} 
\usepackage{epsfig, amssymb}
\def\be{\begin{eqnarray}}
\def\ee{\end{eqnarray}}
\def\bee{\begin{eqnarray*}}
\def\eee{\end{eqnarray*}}
\newtheorem{thm}{Theorem}
\newtheorem{cor}{Corollary}
\newtheorem{lem}{Lemma}

\newtheorem{prop}{Proposition}
\newtheorem{defn}{Definition}

\begin{document}
\title{On local unitary versus local Clifford equivalence of stabilizer states}
\author{Maarten Van den Nest, Jeroen Dehaene, Bart De Moor}
\affiliation{ESAT-SCD, K.U. Leuven, Kasteelpark Arenberg 10, B-3001 Leuven, Belgium
\email{mvandenn@esat.kuleuven.ac.be} }
\date{\today}

%
\begin{abstract}
We study the relation between local unitary (LU) equivalence and local Clifford (LC) equivalence of
stabilizer states. We introduce a large subclass of stabilizer states, such that every two LU
equivalent states in this class are necessarily LC equivalent. Together with earlier results, this
shows that LC, LU and SLOCC equivalence are the same notions for this class of stabilizer states.
Moreover, recognizing whether two given stabilizer states in the present subclass are locally
equivalent only requires a polynomial number of operations in the number of qubits.
\end{abstract}
 \maketitle

\section{Introduction}

Stabilizer states constitute a class of multipartite pure states that play an important role in
numerous tasks in quantum information theory, such as quantum error correction \cite{Gott} and
measurement-based quantum computation \cite{1wayQC}. A stabilizer state on $n$ qubits is defined as
a simultaneous eigenvector of a maximal set of commuting operators in the Pauli group on $n$
qubits, where the latter is the group generated by all $n$-fold tensor products of the Pauli
matrices and the identity.

In order to understand the role of stabilizer states in existing and possibly new applications, the
properties of these states have recently been studied by numerous authors (see e.g.
\cite{entgraphstate, graphbriegel, Bell_graph, val_bond, localcliffgraph, alg_codes, invar_stab,
cliff_inv_prl}). One major open problem is a classification of stabilizer states in local unitary
(LU) equivalence classes: not only is this problem of natural importance in the study of the
entanglement properties of stabilizer states, but it is also relevant both in the quantum coding
aspect of stabilizer states as in their role in the one-way quantum computing model. When studying
LU equivalence of stabilizer states, it is natural to consider, in a first step, only those LU
operations that belong to the local Clifford (LC) group, where the latter consists of all local
unitary operations that map the Pauli group to itself under conjugation. Indeed, the stabilizer
formalism plus the (local) Clifford group form a closed framework which can entirely be described
in terms of binary linear algebra, and this binary description simplifies the study of LC
equivalence of stabilizer states to a great extent with respect to general LU equivalence. We have
studied several aspects of LC equivalence of stabilizer states in earlier work
\cite{localcliffgraph, alg_codes, cliff_inv_prl}. In a second step, it is natural to raise the
question whether the restriction of considering only LC operations is in fact a restriction at all.
In other words, the question is asked whether every two LU equivalent stabilizer states are also LC
equivalent, or, conversely, whether there exist LU equivalent stabilizer states that are not
related by a local Clifford operation. This problem will be our topic of interest in the following.
In particular, we will show that the answer to the above question is positive for a large subclass
of stabilizer states and it is, with the present result, our aim to take the first
steps towards a complete answer to the above question. 

To construct our subclass of stabilizer states, we elaborate on an approach that was used in Ref.
\cite{RainsMin2} to prove that any two LU equivalent so-called GF(4)-linear stabilizer codes are
also LC equivalent (this terminology is discussed below). The main concept introduced in the above
reference to study the problem at hand is that of \emph{minimal support} of a stabilizer. In the
following, we study this notion in more detail and construct an extension of the class of
(self-dual) GF(4)-linear stabilizer codes, such that every two LU equivalent states in this class
must also be LC equivalent.

Finally, we wish to point out that the present result, and a possible general proof of the
assertion that LU and LC equivalence of stabilizer states are identical notions, has several
interesting implications. Firstly, we showed in earlier work \cite{MTNS} that any two stabilizer
states related by an (invertible) stochastic local operation assisted with classical communication
(SLOCC) are also LU equivalent. Together with the present result, this shows that SLOCC, LU and LC
equivalence are identical notions within the present subclass of stabilizer states, and therefore
there is in fact only a single notion of 'local' equivalence within this class of states. Secondly,
in Ref. \cite{alg_codes} we presented an algorithm of polynomial complexity in the number of qubits
which recognizes whether two given stabilizer states are LC equivalent. This shows that all three
local equivalences can be detected efficiently within the present class of stabilizer states.


\section{Stabilizer formalism}

In this section we state the necessary preliminaries concerning the stabilizer formalism and the
local Clifford group.

\subsection{Stabilizer states, LU and LC equivalence}

The Pauli group ${\cal G}_n$ on $n$ qubits consists of all $4\times 4^n$ local operators of the
form $M= \alpha_M M_1\otimes\dots\otimes M_n$, where $\alpha_M\in\{\pm 1, \pm i\}$ is an overal
phase factor and $M_i$ is either the $2\times 2$ identity matrix $\sigma_0$ or one of the Pauli
matrices $\sigma_x$, $\sigma_y$, $\sigma_z$. The Clifford group ${\cal C}_1$ on one qubit is the
group of all $2\times 2$ unitary operators that map $\sigma_u$ to $\alpha_u\sigma_{\pi(u)}$ under
conjugation, where $u=x,y,z$, for some $\alpha_u=\pm 1$ and  some permutation $\pi$ of $\{x,y,z\}$.
The local Clifford group ${\cal C}_n^l$ on $n$ qubits is the $n$-fold tensor product of ${\cal
C}_1$ with itself.

A stabilizer ${\cal S}$ in the Pauli group is defined as an abelian subgroup of ${\cal G}_n$ which
does not contain $-I$ \cite{QCQI}. A stabilizer consists of $2^k$ Hermitian Pauli operators (i.e.
they must have real overall phase factors $\pm 1 $), for some $k\leq n$. As the operators in a
stabilizer commute, they can be diagonalized simultaneously and, what is more, if $|{\cal S}|=2^n$
then there exists a unique state $|\psi\rangle$ on $n$ qubits such that $M|\psi\rangle =
|\psi\rangle$ for every $M\in{\cal S}$. Such a state $|\psi\rangle$ is called a stabilizer state
and the group ${\cal S}={\cal S}(\psi)$ is called the stabilizer of $|\psi\rangle$. The expansion
\be\label{sumstab} |\psi\rangle\langle\psi| = \frac{1}{2^n}\sum_{M\in {\cal S}(\psi)} M, \ee which
describes a stabilizer state as a sum of all elements in its stabilizer, can readily be verified.

The support supp$(M)$ of an element $M = \alpha_M M_1\otimes\dots\otimes M_n\in{\cal S}(\psi)$ is
the set of all $i\in \{1, \dots, n \}$ such that $M_i$ differs from the identity. Let
$\omega=\{i_1, \dots, i_k\}$ be a subset of $\{1, \dots, n\}$. Tracing out all qubits of
$|\psi\rangle$ outside $\omega$ yields a (generally mixed) state $\rho_{\omega}(\psi)$, which is
equal to \be\rho_{\omega}(\psi)=\frac{1}{2^{|\omega|}}\sum_{M\in{\cal S}, \mbox{ \scriptsize
supp}(M)\subseteq\omega} \alpha_M M_{i_1}\otimes\dots\otimes M_{i_k}.\ee  This can easily be
verified using the identity (\ref{sumstab}). We denote by $A_{\omega}(\psi)$ the number of elements
$M\in {\cal S}(\psi)$ with supp$(M)=\omega$. It is important to note that the function
$A_{\omega}(\cdot)$ is an LU invariant, i.e. it takes on equal values on LU equivalent stabilizer
states \cite{cliff_inv_prl}.

Two stabilizer states are called LU (LC) equivalent if there exists a local unitary (local
Clifford) operator which relates these two states. If $|\psi\rangle$ is a stabilizer state, the set
LU$(\psi)$ consists of all stabilizer states that are LU equivalent to $|\psi\rangle$. The set
LC$(\psi)$ is defined analogously.

Finally, a multipartite pure state is called \emph{fully entangled} if it cannot be written as a
tensor product of two states. It is clear that in the present context there is no restriction in
considering only fully entangled stabilizer states, and we will suppose that every stabilizer state
in the following is fully entangled. To avoid technical details that appear when dealing with small
numbers of qubits, we will also only consider stabilizer states on $n\geq 3$ qubits \footnote{We
note that LU$(\psi)=$ LC$(\psi)$ for every stabilizer state $|\psi\rangle$ on $n\leq 2$ qubits;
this can trivially be verified for the 1-qubit case; for the 2-qubit case, this result essentially
follows from the fact that, up to LC equivalence, the only fully entangled stabilizer state is the
EPR state $(|00\rangle+|11\rangle)/\sqrt{2}$.}.

\subsection{Binary representation}

It is well known that the stabilizer formalism  has an equivalent formulation in terms of algebra
over the field $\mathbb{F}_2=$ GF(2), where arithmetic is performed modulo two. The heart of this
binary representation of stabilizers is the mapping
\begin{eqnarray}\label{pairsbits}
\sigma_0=\sigma_{00} &\mapsto& (0,0)\nonumber\\
\sigma_x=\sigma_{01} &\mapsto& (0,1)\nonumber\\
\sigma_z=\sigma_{10} &\mapsto& (1,0)\nonumber\\
\sigma_y=\sigma_{11} &\mapsto& (1,1),
\end{eqnarray}
which encodes the Pauli matrices as pairs of bits. Consequently, the elements of ${\cal G}_n$ can
be represented as $2n$-dimensional binary vectors as follows:
\be\label{2ndim}\sigma_{w_1w_1'}\otimes\dots\otimes\sigma_{w_nw_n'}=\sigma_{(w,w')} \mapsto  (w,w')
\in \Bbb{F}_2^{2n},\ee where $w:=(w_1, \dots, w_n),\ w':=(w_1', \dots, w_n')\in\mathbb{F}_2^{n}$
are $n$-dimensional binary vectors. Note that the information about the $i$th qubit is distributed
over the $i$th components of the vectors $w$ and $w'$. The parameterization (\ref{2ndim})
establishes a group homomorphism between ${\cal G}_n, \cdot$ and $\Bbb{F}_2^{2n},+$ (which
disregards the overall phases of Pauli operators). In this binary representation, two Pauli
operators $\sigma_a$ and $\sigma_b$, where $a, b \in \Bbb{F}_2^{2n}$, commute if and only if $a^T P
b = 0$, where the $2n\times 2n$ matrix \be P = \left[
\begin{array}{cc} 0 & I \\ I& 0 \end{array} \right]\ee
defines a symplectic inner product on $\Bbb{F}_2^{2n}$. Therefore, a stabilizer ${\cal S}(\psi)$
 of an $n$-qubit stabilizer state $|\psi\rangle$ corresponds to an $n$-dimensional linear subspace of $\Bbb{F}_2^{2n}$
which is self-dual with respect to this symplectic inner product, i.e., $a^T P b = 0$ for every $a,
b$ in this space. The binary stabilizer space is usually presented in terms of a $2n\times n$
binary matrix $S$, the columns of which form a basis of this space. The entire binary stabilizer
space ${\cal C}_S$ is the column space of $S$. As ${\cal C}_S$ is its own symplectic dual, a vector
$v\in \Bbb{F}_2^{2n}$ belongs to ${\cal C}_S$ if and only if $S^T P v=0$. Consequently, the
generator matrix $S$ satisfies $S^TPS=0$.

An important subclass of stabilizer states is constituted by the graph states. Graph states are
those stabilizer states that have a generator matrix of the form $[\theta\ I]^T$, where $\theta$ is
the $n\times n$ adjacency matrix of a simple graph on $n$ vertices (see e.g.
\cite{localcliffgraph}). Therefore, a graph state on $n$ qubits is in a one-to-one correspondence
with a graph on $n$ vertices. It is well known that every stabilizer state is LC equivalent to some
(generally non-unique) graph state \cite{stabgraphcode}.

When disregarding the overall phases of the elements in ${\cal G}_1$, it is easy to see that there
exists a one-to-one correspondence between the one-qubit Clifford operations and the 6 possible
invertible linear transformations of $\Bbb{F}_2^{2}$, since each one-qubit Clifford operator
performs one of the 6 possible permutations of the Pauli matrices and leaves the identity fixed.
Local Clifford operations $U\in {\cal C}_n^l$ on $n$ qubits then correspond to nonsingular
$2n\times 2n$ binary matrices $Q$ of the block form
\be Q = \left [ \begin{array}{cc} A&B\\
C&D \end{array}\right],\ee where the $n\times n$ blocks $A, B, C, D$ are diagonal
\cite{localcliffgraph}. We denote the diagonal entries of $A, B, C, D$ by
 $a_i$, $b_i$, $c_i$, $d_i$, respectively. The $n$  submatrices \be Q_i:=\left [
\begin{array}{cc} a_i & b_i
\\ c_i& d_i\end{array} \right ]\in GL(2,\Bbb{F}_2)\ee correspond to the tensor factors of $U$.
We denote the group of all such $Q$ by $C^l_n$. Two $n$-qubit stabilizer states $|\psi\rangle$,
$|\psi'\rangle$  with generator matrices $S$, $S'$, respectively, are therefore LC equivalent if
and only if there exists an operator $Q\in C_n^l$ such that ${\cal C}_{QS}={\cal C}_{S'}$, i.e.,
$Q$ maps the space ${\cal C}_{S}$ to the space ${\cal C}_{S'}$. As these spaces are their own
symplectic duals, this is equivalent to stating that $S^{'T} P QS=0$.

\subsection{GF(4) representation and linearity}

There is also a well known representation of stabilizers in terms of algebra over the field
$\Bbb{F}_4=$ GF(4) \cite{codeGF4}. This is the finite field of 4 elements, which can be written as
$\Bbb{F}_4= \{0,1, \xi, \xi^2\}.$ Addition and multiplication in $\Bbb{F}_4$ satisfy the rules \be
1 + 1 = \xi + \xi = \xi^2 + \xi^2 &=& 0\nonumber\\ 1 + \xi &=& \xi^2. \ee Note that addition in
$\Bbb{F}_4$ is performed modulo 2. Similar to (\ref{pairsbits}), one now uses the encoding
\begin{eqnarray}\sigma_0=&\mapsto& 0,\ \sigma_z \mapsto 1,\ \sigma_x \mapsto \xi,\  \sigma_y
\mapsto \xi^2,
\end{eqnarray}
and consequently Pauli operators on $n$ qubits are represented as $n$-dimentional vectors with
entries in $\Bbb{F}_4$. As in the binary description, the multiplicative structure of ${\cal G}_n$
becomes  the additive structure of $\Bbb{F}_4^n$. Therefore, every stabilizer ${\cal S}(\psi)$ on
$n$ qubits corresponds to an \emph{additive} subset (or \emph{code}) of $\Bbb{F}_4^n$, i.e., the
sum of any two vectors in this set again belongs to this set. Analogous to the binary case, this
additive code is presented in terms of a generator matrix, which now has dimensions $n\times n$.
Moreover, the property that a stabilizer is an abelian group can again be translated into a
self-duality property of the corresponding additive code over GF(4) with respect to a certain inner
product. As the details of this inner product are irrelevant in the following, we will omit them
and the interested reader is referred to Ref. \cite{codeGF4}.

Below we will be interested in those specific stabilizers corresponding to codes over GF(4) that
are closed under scalar multiplication with $\xi$. Such codes are genuine linear subspaces of
$\Bbb{F}_4^n$, as they are closed under taking linear combinations with coefficients in
$\Bbb{F}_4$. They are therefore called GF(4)-linear codes.

\section{Minimal supports}

In this section, we develop the necessary concepts for our study of LU versus LC equivalence of
stabilizer states, and we prove our main result.


Let $|\psi\rangle$ be a stabilizer state on $n$ qubits. A minimal support of ${\cal S}(\psi)$ is a
set $\omega\subseteq\{1, \dots, n\}$ such that there exists an element in ${\cal S}(\psi)$ with
support equal to $\omega$, but there exist no elements with support strictly contained in $\omega$.
An element with minimal support is called a minimal element.  Clearly, if $|\psi'\rangle\in \mbox{
LU}(\psi)$ then $\omega$ is also a minimal support of ${\cal S}(\psi')$: this follows from the fact
that the function $A_{\omega'}(\cdot)$ is an LU invariant, for every $\omega'\subseteq\{1, \dots, n
\}$.

We have the following lemma:

\begin{lem}
Let $|\psi\rangle$ be a stabilizer state and let $\omega$ be a minimal support of ${\cal S}(\psi)$.
Then $A_{\omega}(\psi)$ is equal to 1 or 3 and the latter case can only occur if $|\omega|$ is
even.
\end{lem}
{\it Proof:} (i) If $A_{\omega}(\psi)=1$ then we are done. If $A_{\omega}\geq 2$, let $M,M'\in{\cal
S}(\psi)$ be two different elements with supports equal to $\omega$. These elements must satisfy
$\sigma_0\neq M_i\neq M_i'\neq\sigma_0$ for every $i\in\omega$. Indeed, if this were not the case,
then supp$(MM')$ would be strictly contained in $\omega$, which contradicts the given. It follows
that also supp$(MM')=\omega$ and that the set $\{M_i, M_i', (MM')_i\}$ is equal to $\{\sigma_x,
\sigma_y, \sigma_z\}$ for every $i=1, \dots, n$. Consequently,  $M$, $M'$, and $MM'$ are the only
elements in ${\cal S}(\psi)$ with support equal to $\omega$: indeed, suppose there does exist a
fourth element $N\in {\cal S}(\psi)$ with supp$(N)=\omega$; fixing any $i_0\in\omega$, then either
$M_{i_0}$, $M'_{i_0}$ or $(MM')_{i_0}$ is equal to $N_{i_0}$, say $N_{i_0}=M_{i_0}$; but then
supp$(MN)$ is strictly contained in $\omega$, which leads to a contradiction. This proves the first
part of the lemma. Secondly, $|\omega|$ must be even since $M$ and $M'$ commute. \hfill $\square$

If $\omega$ is a minimal support of ${\cal S}(\psi)$, it follows from the proof of lemma 1 that
$\rho_{\omega}(\psi)$ is, up to an LC operation, one of the following two states :
\be&&\frac{1}{2^{|\omega|}} \left(I_{|\omega|} + \sigma_x^{\otimes |\omega|}\right)\\
&&\frac{1}{2^{|\omega|}}\left( I_{|\omega|} + \sigma_x^{\otimes |\omega|} +
(-1)^{|\omega/2|}\sigma_y^{\otimes |\omega|} + \sigma_z^{\otimes |\omega|} \right),\quad
\label{rains}\ee where $I_{|\omega|}$ denotes the identity operator on $|\omega|$ qubits. We denote
the state (\ref{rains}) by $\rho^{[|\omega|, |\omega|-2, 2]}$. The following property of
$\rho^{[|\omega|, |\omega|-2, 2]}$ will be a central ingredient to the proof of our main result
below:

\begin{lem} \cite{RainsMin2}
 Let $m\in \Bbb{N}_0$, $m\geq 2$. Let $\rho, \rho'$ be two (mixed) states on $2m$ qubits, both LC equivalent to
 $\rho^{[2m,2m-2, 2]}$, and let $U\in U(2)^{\otimes 2m}$ be a local unitary operator such that $U \rho
U^{\dagger}=\rho'.$ Then $U\in {\cal C}_n^l$.
\end{lem}

Lemma 2 is in fact a variant of a result in Ref. \cite{RainsMin2} and the reader is referred to
this reference for a proof.

{\it Remark.} If $m=1$ then \be \rho^{[2,0, 2]} = \frac{1}{4}(I + \sigma_x^{\otimes 2} -
\sigma_y^{\otimes 2}+ \sigma_z^{\otimes 2})\ee is the rank one projection operator associated with
the EPR state  $(|00\rangle+|11\rangle)/\sqrt{2}$ (which is a stabilizer state on 2 qubits).

We will now use the concept of minimal support to construct our class of stabilizer states
$|\psi\rangle$ that satisfy LU$(\psi)=\mbox{ LC}(\psi)$. In other words, every two LU equivalent
states in this class are necessarily LC equivalent. Denoting by ${\cal M}(\psi)$ the subgroup of
${\cal S}(\psi)$ generated by all minimal elements, we are ready to state the central result of
this paper:

\begin{thm}
Let $|\psi\rangle$ be a fully entangled $n$-qubit stabilizer such that $\sigma_x$, $\sigma_y$,
$\sigma_z$ occur on every qubit in ${\cal M}(\psi)$. Then $LU(\psi)=LC(\psi)$.
\end{thm}
{\it Proof: } Let $|\psi'\rangle\in \mbox{ LU}(\psi)$ and fix $U=U_1\otimes\dots\otimes U_n\in
U(2)^{\otimes n}$ such that $U|\psi\rangle= |\psi'\rangle$. We will show that $U_i$ is a Clifford
operation, for every $i=1, \dots, n$. Considering e.g. the first qubit, there exists a minimal
element $M = \alpha_M M_1\otimes\dots\otimes M_n\in {\cal S}(\psi)$ such that $M_1\neq\sigma_0$,
say  $M_1=\sigma_x$. Let $\omega=\{1=i_1, i_2, \dots, i_k \}\subseteq\{1, \dots, n\}$, where
$k=|\omega|$, denote the support of $M$. Then $A_{\omega}(\psi)$ is equal to either 1 or 3 from
lemma 1. We will make a distinction between these two cases.

Firstly, suppose that $A_{\omega}(\psi)=3$. Then $\rho_{\omega}(\psi)$ is LC equivalent to
$\rho^{[|\omega|, |\omega|-2, 2]}$ from the above.  Moreover, as $|\psi'\rangle\in \mbox{
LU}(\psi)$, the set $\omega$ is also a minimal support of ${\cal S}(\psi')$  with
$A_{\omega}(\psi')=A_{\omega}(\psi)=3$. Therefore, $\rho_{\omega}(\psi')$ is LC equivalent to
$\rho^{[|\omega|, |\omega|-2, 2]}$ as well. Using the notation $U_{\omega}=
U_{i_1}\otimes\dots\otimes U_{i_k}$, it follows from $U|\psi\rangle=|\psi'\rangle$ that
$U_{\omega}$ maps $\rho_{\omega}(\psi)$ to $\rho_{\omega}(\psi')$ under conjugation. Now, note that
we must have $|\omega|> 2$; indeed, if $|\omega|$ were equal to 2 then it follows from the remark
below lemma 2 that $\rho_{\omega}(\psi)$ is a pure state; this is however impossible as
$|\psi\rangle$ is fully entangled. Moreover, $|\omega|$ is even from lemma 1, and we can therefore
conclude that $|\omega|\geq 4$. We can now use lemma 2, finding that $U_{\omega}\in {\cal
C}_{k}^l$, and, in particular, $U_1$ is a Clifford operation.

Secondly, let $A_{\omega}(\psi)=1$. Then there exists another minimal element $N\in {\cal M}(\psi)$
such that $1\in\mbox{ supp}(N)$ and $M_1\neq N_1$ from the assumption in the theorem, say $
N_1=\sigma_z$. Let $\mu=\mbox{ supp}(N)$. Note that if $A_{\mu}(\psi)=3$, we can apply the same
argument as above and conclude that $U_1$ is a Clifford operation. On the other hand, suppose that
$A_{\mu}(\psi)=1$. Denoting $M_{\omega} = \alpha_M M_{i_1}\otimes\dots\otimes M_{i_k}$, and
$N_{\mu}$ analogously, we can write
\be \rho_{\omega}(\psi) = \frac{1}{2^{|\omega|}}\left( I_{|\omega|} + M_{\omega}\right)\nonumber\\
\rho_{\mu}(\psi) = \frac{1}{2^{|\mu|}}\left( I_{|\mu|} + N_{\mu}\right).\label{thm1} \ee Moreover,
since the sets $\omega$ and $\mu$ are also minimal supports of ${\cal S}(\psi')$ with
$A_{\mu}(\psi')=A_{\omega}(\psi')=1$, there exist unique $M', N'\in {\cal S}(\psi')$
such that \be \rho_{\omega}(\psi') = \frac{1}{2^{|\omega|}}\left( I_{|\omega|} + M_{\omega}'\right)\nonumber\\
\rho_{\mu}(\psi') = \frac{1}{2^{|\mu|}}\left( I_{|\mu|} + N_{\mu}'\right),\label{thm1'} \ee where
$M_{\omega}'$ and $N_{\mu}'$ are defined analogously to $M_{\omega}$ and $N_{\mu}$. Now, as
\be U_{\omega}\rho_{\omega}(\psi)U_{\omega}^{\dagger}&=&\rho_{\omega}(\psi')\nonumber\\
U_{\mu}\rho_{\mu}(\psi)U_{\mu}^{\dagger}&=&\rho_{\mu}(\psi'),\ee where again we have used the
notation $U_{\omega}= U_{i_1}\otimes\dots\otimes U_{i_k}$ and analogously for $U_{\mu}$, we have
\be U_1 M_1 U_1^{\dagger} &=& \pm M_1'\nonumber\\
U_1 N_1 U_1^{\dagger} &=& \pm N_1'\label{thm1''}\ee from (\ref{thm1})-(\ref{thm1'}). Using
$M_1=\sigma_x$ and $N_1=\sigma_z$,  The identities (\ref{thm1''}) show that $U_1$ is a Clifford
operation.

Repeating the above arguments for all $n$ qubits yields the result. \hfill $\square$

Theorem 1 shows that it is sufficient that the group ${\cal M}(\psi)$ has a sufficiently rich
structure in order for LU$(\psi)$ to be equal to LC$(\psi)$. To gain insight in which states meet
the requirement of theorem 1, it is instructive to consider some sufficient conditions for this
criterion to hold. Several sufficient conditions are summarized in corollary 1. We note that case
(ii) of corollary 1 has already been proved in Ref. \cite{RainsMin2}.

\begin{cor}
Let $|\psi\rangle$ be a stabilizer state on $n$-qubits such that one of the following assertions
(i)-(iv) is true. Then $LU(\psi)=LC(\psi)$.

(i) ${\cal S}(\psi)={\cal M}(\psi)$, i.e. ${\cal S}(\psi)$ is generated by its minimal elements.

(ii) The stabilizer ${\cal S}(\psi)$ corresponds to a $GF(4)$-linear code.

(iii) For every $M\in{\cal S}(\psi)$ with nonminimal support, there exists a minimal support
$\omega'\subset\ supp(M)$ such that $A_{\omega'}(\psi)=3$.

(iv) There exists minimal supports $\omega_1, \omega_2, \dots$ satisfying $3= A_{\omega_1}(\psi)=
A_{\omega_2}(\psi)=\dots$, such that every $i\in\{1, \dots, n\}$ belongs to at least one
$\omega_j$.
\end{cor}

{\it Proof:} We will show that (i)-(iv) imply that $\sigma_x$, $\sigma_y$ and $\sigma_z$ occur on
every qubit in ${\cal M}(\psi)$.

Firstly, using lemma 3 (stated below), one  immediately finds that assertion (i) implies the
desired result.

Secondly, suppose (ii) holds. To prove the result, note that every linear subspace of $GF(q)^n$,
with $q$ any prime power, is generated by its minimal elements \cite{min_vect} (here, the notions
of (minimal) support and minimal element are defined in the natural way). Using (i) then yields the
result.

Thirdly, we show that (iii) also implies (i): let $M\in {\cal S}(\psi)$ be an arbitrary nonzero
stabilizer element. We have to show that $M$ is a product of minimal elements. If $M$ is minimal
then we are done. For nonminimal $M$, we will prove the assertion by induction on
$|\mbox{supp}(M)|$. Let $\omega'$ be a minimal support such that $\omega'\subset\mbox{ supp}(M)$
and $A_{\omega'}(\psi)=3$. As $A_{\omega'}(\psi)=3$, there exists a minimal codeword $M'\in {\cal
M}(\psi)$ with support equal to $\omega'$ such that $M$ and $M'$ are equal on the first qubit (see
proof of lemma 1). Consequently, $|\mbox{supp}(MM')|$ is strictly smaller than $|\mbox{supp}(M)|$
and therefore $MM'\in {\cal M}(\psi)$ by induction. But then also $M\in {\cal M}(\psi)$, since $M'$
is minimal. This shows that (iii) implies (i).

Fourthly, suppose that assertion (iv) holds. It then immediately follows from the proof of lemma 1
that $\sigma_x$, $\sigma_y$ and $\sigma_z$ occur on every qubit in ${\cal M}(\psi)$.  \hfill
$\square$

\begin{lem}
Let $|\psi\rangle$ be a (fully entangled) stabilizer state on $n\geq 2$ qubits. Then all three
Pauli matrices $\sigma_x$, $\sigma_y$, $\sigma_z$ occur on every qubit in ${\cal S}(\psi)$.
\end{lem}

The proof of lemma 3 is given in appendix A.

Clearly, conditions (iii) and (iv) are are much more operational than theorem 1, as it is
sufficient to know (only part of) the list of invariants $A_{\omega}(\psi)$ of a given state
$|\psi\rangle$ in order to conclude that LU$(\psi)$ is equal to LC$(\psi)$. Also condition (ii) is
easy to check: indeed, we will show in the next section that one can characterize stabilizers that
correspond to GF(4)-linear codes through a very simple constraint on their binary generator matrix.

\section{Example: GF(4)-linear codes}

In this section we give a simple characterization of those stabilizers ${\cal S}(\psi)$ that
correspond to $GF(4)$-linear stabilizer spaces.

It follows from the discussion in section II that every element in $ \mathbb{F}_4$ has the form $a1
+ b\xi \equiv a + b\xi $, where $(a, b)\in \mathbb{F}_2^2$. Using the identity $\xi^2 = \xi + 1$,
multiplication of $a + b\xi$ with a second element $a' + b'\xi $, where also $(a', b')\in
\mathbb{F}_2^2$, yields \be (a + b\xi)(a' + b'\xi) = aa' + bb' + (ab' + ba' + bb')\xi.\ee
Consequently, the set $\mathbb{F}_2^2$ inherits a multiplication law $*$ from ${\mathbb{F}}_4$,
defined by \be \label{mult}(a, b)*(a',b')= (aa'+bb', ab' + ba' + bb'),\ee and $\mathbb{F}_2^2$,
 with the multiplication $*$ and the standard addition modulo 2, is a field isomorphic to
${\mathbb{F}}_4$.
 It follows that the set $\mathbb{F}_2^{2n}$ with the standard addition of vectors modulo
2 and the scalar multiplication $*$, is an $n$-dimensional vector space over $\Bbb{F}_4$; here, the
scalar multiplication is defined as follows: letting $v\in \mathbb{F}_2^{2n}$ and $(a, b)\in
\mathbb{F}_2^2$, the vector $w:=(a, b)*v$ is defined by \be (w_i, w_{n+i}) = (a,b)*(v_i,
v_{n+i}),\ee for every $i=1, \dots, n$. With these definitions, it follows from the discussion in
section II that a binary stabilizer space ${\cal C}_S$ is
 GF(4)-linear if and only if $(0, 1)* v \in {\cal C}_S$ for every $v\in {\cal C}_S$. Note that the action  $(a,b)\to
(0, 1)*(a,b)$ is linear transformation on the vector $(a,b)\in \mathbb{F}_2^2$. Indeed, one has
\be\label{lin_action} (0, 1)*(a,b) = \left[\begin{array}{cc} 0&1\\1&1\end{array}\right]
\left[\begin{array}{c} a\\b\end{array}\right].\ee Consequently, if $v\in\mathbb{F}_2^{2n}$ then
$(0, 1)*v$ corresponds to \be\label{F_4'}\left[\begin{array}{cc} 0 &I\\I&I\end{array}\right] v,\ee
where $I$ is the $n\times n$ identity matrix and $0$ is here the $n\times n$ zero matrix. This
leads to a simple characterization of $GF(4)$-linear stabilizer spaces: let ${\cal C}_S$ be a
binary stabilizer space with generator matrix $S$. Denoting the columns of $S$ by $s^j$ ($j=1,
\dots, n$), the space ${\cal C}_S$ is $GF(4)$-linear if and only if \be (0, 1)*s^j\in {\cal C}_S\ee
for every $j=1, \dots, n$, as the columns of $S$ form a basis of ${\cal C}_S$ (regarded as a vector
space over $\Bbb{F}_2$). Using (\ref{F_4'}) and the fact that any vector $v\in \Bbb{F}_2^{2n}$
belongs to ${\cal C}_S$ if and only if $S^T P v=0$, we find that ${\cal C}_S$ is $GF(4)$-linear if
and only if \be\label{F_4''} S^T P \left[\begin{array}{cc} 0&I\\I&I\end{array}\right]S = 0.\ee
Denoting $S = [S_z^T\ S_x^T]^T$, where $S_z$, $S_x$ are $n\times n$ blocks, (\ref{F_4''}) is
equivalent to \be 0 &=& S_z^T S_z + S_x^T S_x + S_z^T S_x.\ee We have proven the following theorem:

\begin{thm}
Let $|\psi\rangle$ be a stabilizer state on $n$ qubits with
generator matrix $S= [S_z^T\ S_x^T]^T$. Then ${\cal C}_S$ is
$GF(4)$-linear if and only if \be S_z^T S_z + S_x^T S_x + S_z^T
S_x=0.\ee
\end{thm}

Theorem 2 shows that it is indeed easy to check whether a stabilizer corresponds to a GF(4)-linear
code. It is interesting to note that the $2n\times 2n$ matrix
\be\label{CliffGF4}\left[\begin{array}{cc} 0&I\\I&I\end{array}\right]\ee  in (\ref{F_4''}) belongs
to the group $C_n^l$ and is therefore the binary representation of a local Clifford operation $V\in
{\cal C}_n^l$ \footnote{To be precise, it is the binary representation of a set of local Clifford
operations, since a local Clifford operation is defined by an element in $C_n^l$ \emph{and} a set
of $2n$ phases $\pm 1$.}. What is more, every tensor factor of $V$ corresponds to the same binary
operator, namely the $2\times 2$ matrix which appears in (\ref{lin_action}), which in turn
corresponds to a cyclic permutation $\sigma_x \to\sigma_y\to\sigma_z\to\sigma_x$ of the three Pauli
matrices. Thus, (\ref{F_4''}) expresses that the operator (\ref{CliffGF4}) maps the space ${\cal
C}_S$ to itself or, equivalently, that $V|\psi\rangle = |\psi\rangle$.

\section{Discussion and conclusion}

We will now discuss the results in this paper. Our main result was the presentation of a subclass
of stabilizer states, such that any two LU equivalent stabilizer states in this class must also be
LC
equivalent. 
To construct our class, a central concept was that of minimal support. In particular, we showed
that if the subgroup ${\cal M}(\psi)$ generated by all minimal elements in a stabilizer ${\cal
S}(\psi)$ has a sufficiently rich structure, then the LU equivalence class and the LC equivalence
class  of the state $|\psi\rangle$ coincide.

The main objectives of this work were to give support to the conjecture that LU$(\psi)=\mbox{
LC}(\psi)$ for every stabilizer state and to investigate the relevance of the notion of minimal
support in this problem. The next insight that needs to be gained in the present research, is what
constraints are imposed on those stabilizer states that do not meet the requirement of theorem 1
and what the structure and the size of this remaining set of states is. It is in fact not easy to
find many examples of states outside of our class, and it is not unlikely that the constraints
imposed on such states are sufficiently strict, such that the question whether LU$(\psi)=\mbox{
LC}(\psi)$ for these remaining states can be settled by considering these states case by case. We
note that it is not our hope that our class covers all stabilizer states. Indeed, there do exist
states that do not belong to the class we have considered. An important example is the generalized
GHZ state $|GHZ_n\rangle$ on $n$ qubits, which has a stabilizer ${\cal S }(GHZ_n)$ generated by the
elements \be\label{GHZ}
&&\sigma_x\otimes\sigma_x\otimes\sigma_x\otimes\sigma_x\otimes\dots\otimes\sigma_x,\nonumber\\
&&\sigma_z\otimes\sigma_z\otimes\sigma_0\otimes\sigma_0\otimes\dots\otimes\sigma_0, \nonumber\\
&&\sigma_0\otimes\sigma_z\otimes\sigma_z\otimes\sigma_0\otimes\dots\otimes\sigma_0, \nonumber\\
&&\sigma_0\otimes\sigma_0\otimes\sigma_z\otimes\sigma_z\otimes\dots\otimes\sigma_0,\nonumber\\
&&\dots\ee  The subgroup ${\cal M}(GHZ_n)$ consists of all possible elements $M=
M_1\otimes\dots\otimes M_n$ with $M_i\in\{\sigma_0, \sigma_z\}$, for every $i=1, \dots, n$, and
therefore $|GHZ_n\rangle$ does not meet the requirement of theorem 1. However, also for the GHZ
state (and therefore for every state in its LC equivalence class) one has LU$(GHZ_n)=\mbox{
LC}(GHZ_n)$. This is stated in the following proposition, which is proven in appendix B.

\begin{prop}
Let $|\psi\rangle$ be a fully entangled stabilizer state. Then $|\psi\rangle\in LC(GHZ_n)$ if and
only if $A_{\omega}(\psi)=1$ for every $\omega\in\{\{1, 2\}, \{2, 3\}, \dots\}$. Consequently,
LU$(GHZ_n)=\mbox{ LC}(GHZ_n)$.
\end{prop}

Therefore, while the GHZ state does not belong to our subclass of stabilizer states, it does still
satisfy LU$(GHZ_n)=\mbox{ LC}(GHZ_n)$, which can be proven with simple arguments. It is to date not
clear if there exist other stabilizer states that do not satisfy the conditions of theorem 1 and
hence do not belong to our subclass.

In conclusion, we have studied the relation between local unitary equivalence and local Clifford
equivalence of stabilizer states. In particular, we have investigated the question whether every
two LU equivalent stabilizer states are also LC equivalent. We have shown that the answer to this
question is positive for a large class of stabilizer states, which can be regarded as an extension
of the set of those stabilizer states that correspond to GF(4)-linear codes. We have given
sufficient conditions for a state to belong to our specific class and, in particular, we have given
a simple characterization of stabilizer states corresponding to GF(4)-linear codes in terms of the
binary stabilizer formalism.

\begin{acknowledgments}

MVDN thanks H. Briegel for interesting discussions concerning stabilizer states and for inviting
him to the Techn. Un. Innsbr\"uck; MVDN thanks M. Hein, J. Eisert and D. Schlingemann for
interesting discussions concerning LU and LC equivalence of stabilizer states. The authors thank
the referee for a thorough reading of the manuscript. This research is supported by several funding
agencies: Research Council KUL: GOA-Mefisto 666, GOA-Ambiorics, several PhD/postdoc and fellow
grants; Flemish Government: - FWO: PhD/postdoc grants, projects, G.0240.99 (multilinear algebra),
G.0407.02 (support vector machines), G.0197.02 (power islands), G.0141.03 (Identification and
cryptography), G.0491.03 (control for intensive care glycemia), G.0120.03 (QIT), G.0452.04 (QC),
G.0499.04 (robust SVM), research communities (ICCoS, ANMMM, MLDM); -   AWI: Bil. Int. Collaboration
Hungary/ Poland; - IWT: PhD Grants, GBOU (McKnow) Belgian Federal Government: Belgian Federal
Science Policy Office: IUAP V-22 (Dynamical Systems and Control: Computation, Identification and
Modelling, 2002-2006), PODO-II (CP/01/40: TMS and Sustainibility); EU: FP5-Quprodis;  ERNSI; Eureka
2063-IMPACT; Eureka 2419-FliTE; Contract Research/agreements: ISMC/IPCOS, Data4s, TML, Elia, LMS,
IPCOS, Mastercard; QUIPROCONE; QUPRODIS.
\end{acknowledgments}

\appendix
\section{Proof of lemma 3}

Let $|\psi\rangle$ be a (fully entangled) stabilizer state on $n\geq 2$ qubits. Consider e.g. the
first qubit. Denote $\omega=\{2, \dots, n\}$ and let ${\cal S}_{\omega}$ be the set consisting of
all $M\in {\cal S}(\psi)$ such that $M_1=\sigma_0$, which is a subgroup of ${\cal S}(\psi)$.
Firstly, note that it is impossible that ${\cal S}_{\omega} = {\cal S}(\psi)$; indeed, if this were
the case, then the set \be\label{app}\label{xyz} {\cal S}_{\omega}^{\times}=\left\{M_{\omega}\ |\
M\in{\cal S}(\psi)\right\}\ee would be a stabilizer on $n-1$ qubits with cardinality $2^n$, which
is a contradiction (in (\ref{app}), we have used the notation $M_{\omega}= \alpha_M
M_2\otimes\dots\otimes M_n$ as before). Secondly, suppose that there exists an $a\in\{x, y, z\}$
such that $M_1\in\{\sigma_0, \sigma_a\}$ for every $M\in{\cal S}(\psi)$ (and both $\sigma_0$ and
$\sigma_a$ occur). It is then easy to verify that \be\label{coset}{\cal S}(\psi) = {\cal
S}_{\omega}\cup M^a {\cal S}_{\omega}\ee where $M^a$ is an arbitrary element in ${\cal S}(\psi)$
satisfying $M^a_1 = \sigma_a$, and $M^a {\cal S}_{\omega}$ is the coset of ${\cal S}_{\omega}$
determined by the element $M^a$. Therefore, $|{\cal S}_{\omega}|=2^{n-1}$ and the stabilizer
(\ref{xyz}) now has cardinality $2^{n-1}$, consequently defining a stabilizer state
$|\psi_{\omega}\rangle$ on $n-1$ qubits. Using the identities \be |\psi\rangle\langle\psi| &=&
\frac{1}{2^n} \sum_{M\in {\cal S}} M \nonumber\\ |\psi_{\omega}\rangle\langle\psi_{\omega}| &=&
\frac{1}{2^{n-1}} \sum_{N\in {\cal S}_{\omega}^{\times}} N\ee and (\ref{coset}), it easily follows
that \be|\psi\rangle\langle\psi| =\frac{1}{2}
(I+M^a)\left(\sigma_0\otimes|\psi_{\omega}\rangle\langle\psi_{\omega}|\right)\ee and therefore
$\mbox{Tr}_1\left\{ |\psi\rangle\langle\psi|\right\}= |\psi_{\omega}\rangle\langle\psi_{\omega}|.$
This shows that $\mbox{Tr}_1|\psi\rangle\langle\psi|$ is a pure state, which leads to a
contradiction, as $|\psi\rangle$ is fully entangled. It follows that $\sigma_x$, $\sigma_y$,
$\sigma_z$ must occur on the first qubit of ${\cal S}(\psi)$. Repeating the above argument for all
qubits yields the result. \hfill $\square$

\section{Proof of proposition 1}

If $|\psi\rangle\in \mbox{ LU}(GHZ_n)$ then clearly $A_{\omega}(\psi)=1$ for every $\omega\in\{\{1,
2\}, \{2, 3\}, \dots\}$. Conversely, suppose that $|\psi\rangle$ is a fully entangled stabilizer
state on $n$ qubits such that $A_{\omega}(\psi)=1$ for every $\omega\in\{\{1, 2\}, \{2, 3\},
\dots\}$. Then ${\cal S}(\psi)$ has $n-1$ elements \be
&&\alpha_1\ \sigma_{k_1}\otimes\sigma_{k_2}\otimes\sigma_0\otimes\sigma_0\otimes\dots\otimes\sigma_0, \nonumber\\
&&\alpha_2\ \sigma_0\otimes\sigma_{l_2}\otimes\sigma_{l_3}\otimes\sigma_0\otimes\dots\otimes\sigma_0, \nonumber\\
&&\alpha_3\ \sigma_0\otimes\sigma_0\otimes\sigma_{m_3}\otimes\sigma_{m_4}\otimes\dots\otimes\sigma_0,\nonumber\\
&&\dots\ee where $\alpha_1, \alpha_2, \alpha_3, \dots \in \{\pm 1\}$ and $k_1, k_2, l_2,l_3, \dots
\in\{x, y, z\}$. Moreover, $k_2=l_2$, $l_3=m_3, \dots$ since ${\cal S}(\psi)$ is an abelian group .
This yields $n-1$ independent \footnote{By "independent" is meant that no element in this set can
be written as a product of the others.} elements of ${\cal S}(\psi)$. Therefore, to obtain a
complete set of $n$ generators, we need one additional element. We claim that any other generator
$M=\alpha_n M_1\otimes\dots\otimes M_n$ of ${\cal S}(\psi)$ must have full support and additionally
satisfy $M_1\neq \sigma_{k_1}$, $M_2\neq \sigma_{l_2}, \dots$: indeed, this readily follows from
lemma 3. We therefore obtain a set of generators of the form \be
&&\gamma_1\ \sigma_{a_1}\otimes\sigma_{a_2}\otimes\sigma_{a_3}\otimes\sigma_{a_4}\otimes\dots\otimes\sigma_{a_n},\nonumber\\
&&\gamma_2\ \sigma_{b_1}\otimes\sigma_{b_2}\otimes\sigma_0\otimes\sigma_0\otimes\dots\otimes\sigma_0, \nonumber\\
&&\gamma_3\ \sigma_0\otimes\sigma_{b_2}\otimes\sigma_{b_3}\otimes\sigma_0\otimes\dots\otimes\sigma_0, \nonumber\\
&&\gamma_4\ \sigma_0\otimes\sigma_0\otimes\sigma_{b_3}\otimes\sigma_{b_4}\otimes\dots\otimes\sigma_0,\nonumber\\
&&\dots, \ee with $\gamma_i\in\{\pm 1\}$,  $a_i, b_i\in \{x, y, z\}$ and $a_i\neq b_i$ for every
$i=1, \dots, n$. One can now always find a local Clifford operator which maps these operators to
the set (\ref{GHZ}), which shows that $|\psi\rangle\in LC(GHZ_n)$. Finally, suppose that
$|\psi\rangle$ is a stabilizer state LU equivalent to $|GHZ_n\rangle$. Then $|\psi\rangle$ is fully
entangled and $A_{\omega}(\psi)= A_{\omega}(GHZ)=1$ for every $\omega\in\{\{1, 2\}, \{2, 3\},
\dots\}$. But then $|\psi\rangle\in LC(GHZ_n)$ from the above. This ends the proof. \hfill
$\square$

\bibliographystyle{unsrt}
\bibliography{LU_LC}


\vspace{5cm}

\end{document}